\documentstyle[12pt,twoside]{article}

\def\a{\alpha}
\def\b{\beta}
\def\c{\chi}
\def\d{\delta}
\def\e{\epsilon}                
\def\f{\phi}                    
\def\g{\gamma}
\def\h{\eta}

\def\j{\psi}
\def\k{\kappa}
\def\l{\lambda}
\def\m{\mu}
\def\n{\nu}
\def\o{\omega}
\def\p{\pi}                     
\def\r{\rho}                    

\def\x{\xi}

\def\F{\Phi}

\def\L{\Lambda}

\def\S{\Sigma}

\def\cf{{\cal F}}

\def\ch{{\cal H}}

\def\cl{{\cal L}}

\def\cn{{\cal N}}

\catcode`@=11

\def\un#1{\relax\ifmmode\@@underline#1\else $\@@underline{\hbox{#1}}$\relax\fi}

{\catcode`\'=\active \gdef'{{}~\bgroup\prim@s}}

\def\magstep#1{\ifcase#1 \@m\or 1200\or 1440\or 1728\or 2074\or 2488\or
        2986\fi\relax}

\font\twfvmi=cmmi10\@magscale5
    \skewchar\twfvmi='177
\font\twfvsy=cmsy10\@magscale5
    \skewchar\twfvsy='60
\font\twfvly=lasy10\@magscale5
\font\thtyrm=cmr10\@magscale6

\def\vpt{\textfont\z@\fivrm
  \scriptfont\z@\fivrm \scriptscriptfont\z@\fivrm
\textfont\@ne\fivmi \scriptfont\@ne\fivmi \scriptscriptfont\@ne\fivmi
\textfont\tw@\fivsy \scriptfont\tw@\fivsy \scriptscriptfont\tw@\fivsy
\textfont\thr@@\tenex \scriptfont\thr@@\tenex \scriptscriptfont\thr@@\tenex
\def\prm{\fam\z@\fivrm}%
\def\unboldmath{\everymath{}\everydisplay{}\@nomath
  \unboldmath\fam\@ne\@boldfalse}\@boldfalse
\def\boldmath{\@subfont\boldmath\unboldmath}%
\def\pit{\@getfont\pit\itfam\@vpt{cmti5}}%
\def\psl{\@subfont\sl\it}%
\def\pbf{\@getfont\pbf\bffam\@vpt{cmbx5}}%
\def\ptt{\@subfont\tt\rm}%
\def\psf{\@subfont\sf\rm}%
\def\psc{\@subfont\sc\rm}%
\def\ly{\fam\lyfam\fivly}\textfont\lyfam\fivly
    \scriptfont\lyfam\fivly \scriptscriptfont\lyfam\fivly
\@setstrut\rm}

\def\@vpt{}

\def\vipt{\textfont\z@\sixrm
  \scriptfont\z@\sixrm \scriptscriptfont\z@\sixrm
\textfont\@ne\sixmi \scriptfont\@ne\sixmi \scriptscriptfont\@ne\sixmi
\textfont\tw@\sixsy \scriptfont\tw@\sixsy \scriptscriptfont\tw@\sixsy
\textfont\thr@@\tenex \scriptfont\thr@@\tenex \scriptscriptfont\thr@@\tenex
\def\prm{\fam\z@\sixrm}%
\def\unboldmath{\everymath{}\everydisplay{}\@nomath
  \unboldmath\@boldfalse}\@boldfalse
\def\boldmath{\@subfont\boldmath\unboldmath}%
\def\pit{\@subfont\it\rm}%
\def\psl{\@subfont\sl\rm}%
\def\pbf{\@getfont\pbf\bffam\@vipt{cmbx6}}%
\def\ptt{\@subfont\tt\rm}%
\def\psf{\@subfont\sf\rm}%
\def\psc{\@subfont\sc\rm}%
\def\ly{\fam\lyfam\sixly}\textfont\lyfam\sixly
    \scriptfont\lyfam\sixly \scriptscriptfont\lyfam\sixly
\@setstrut\rm}

\def\@vipt{}

\def\xxxpt{\textfont\z@\thtyrm
  \scriptfont\z@\twfvrm \scriptscriptfont\z@\twtyrm
\textfont\@ne\twfvmi \scriptfont\@ne\twfvmi \scriptscriptfont\@ne\twtymi
\textfont\tw@\twfvsy \scriptfont\tw@\twfvsy \scriptscriptfont\tw@\twtysy
\textfont\thr@@\tenex \scriptfont\thr@@\tenex \scriptscriptfont\thr@@\tenex
\def\unboldmath{\everymath{}\everydisplay{}\@nomath\unboldmath
        \textfont\@ne\twfvmi \textfont\tw@\twfvsy \textfont\lyfam\twfvly
        \@boldfalse}\@boldfalse
\def\boldmath{\@subfont\boldmath\unboldmath}%
\def\prm{\fam\z@\thtyrm}%
\def\pit{\@subfont\it\rm}%
\def\psl{\@subfont\sl\rm}%
\def\pbf{\@getfont\pbf\bffam\@xxxpt{cmbx10\@magscale6}}%
\def\ptt{\@subfont\tt\rm}%
\def\psf{\@subfont\sf\rm}%
\def\psc{\@subfont\sc\rm}%
\def\ly{\fam\lyfam\twfvly}\textfont\lyfam\twfvly
   \scriptfont\lyfam\twfvly \scriptscriptfont\lyfam\twtyly
\@setstrut \rm}

\def\@xxxpt{}

\def\Huge{\@setsize\Huge{36pt}\xxxpt\@xxxpt}

\font\thtymi=cmmi10\@magscale6
    \skewchar\thtymi='177
\font\thtysy=cmsy10\@magscale6
    \skewchar\thtysy='60
\font\thtyly=lasy10\@magscale6
\font\thsirm=cmr12\@magscale6

\def\xxxvipt{\textfont\z@\thsirm
  \scriptfont\z@\thtyrm \scriptscriptfont\z@\twfvrm
\textfont\@ne\thtymi \scriptfont\@ne\thtymi \scriptscriptfont\@ne\twfvmi
\textfont\tw@\thtysy \scriptfont\tw@\thtysy \scriptscriptfont\tw@\twfvsy
\textfont\thr@@\tenex \scriptfont\thr@@\tenex \scriptscriptfont\thr@@\tenex
\def\unboldmath{\everymath{}\everydisplay{}\@nomath\unboldmath
        \textfont\@ne\thtymi \textfont\tw@\thtysy \textfont\lyfam\thtyly
        \@boldfalse}\@boldfalse
\def\boldmath{\@subfont\boldmath\unboldmath}%
\def\prm{\fam\z@\thsirm}%
\def\pit{\@subfont\it\rm}%
\def\psl{\@subfont\sl\rm}%
\def\pbf{\@getfont\pbf\bffam\@xxxpt{cmss12\@magscale6}}%
\def\ptt{\@subfont\tt\rm}%
\def\psf{\@subfont\sf\rm}%
\def\psc{\@subfont\sc\rm}%
\def\ly{\fam\lyfam\thtyly}\textfont\lyfam\thtyly
   \scriptfont\lyfam\thtyly \scriptscriptfont\lyfam\twfvly
\@setstrut \rm}

\def\@xxxvipt{}

\def\HUGE{\@setsize\HUGE{43pt}\xxxvipt\@xxxvipt}

\catcode`@=12

\font\tenex=cmex10 scaled 1200

\def\Sc#1{\hbox{\sc #1}}        

\def\bo{{\raise.05ex\hbox{\large$\Box$}\:}}             
\def\cbo{{\,\raise-.15ex\Sc [\,}}                       
\def\pa{\partial}                                       
\def\su{\sum}                                           
\def\TH{{\raise.2ex\hbox{$\displaystyle \bigodot$}\mskip-4.7mu \llap H \;}}
\def\face{\hbox{\normalsize$\;\;\:{\raise.9ex\hbox{\oo n}\mskip-13mu \llap
        {${\buildrel{\hbox{\frtnrm ..}}\over\smile}$}}\:$}}     
\def\Face{{\raise.2ex\hbox{$\displaystyle \bigodot$}\mskip-2.2mu \llap {$\ddot
        \smile$}}}                                      
\def\Lhat{{\bf\rlap{\kern-.09em$\hat{\phantom L}$}L}}
\def\Lcheck{{\bf\rlap{\kern-.09em$\check{\phantom L}$}L}}

\def\sp#1{{}^{#1}}                              
\def\sb#1{{}_{#1}}                              
\def\sl#1{\rlap{\hbox{$\mskip 1 mu /$}}#1}      
\def\leftrightarrowfill{$\mathsurround=0pt \mathord\leftarrow \mkern-6mu
        \cleaders\hbox{$\mkern-2mu \mathord- \mkern-2mu$}\hfill
        \mkern-6mu \mathord\rightarrow$}
\def\dvec#1{\vbox{\ialign{##\crcr
        \leftrightarrowfill\crcr\noalign{\kern-1pt\nointerlineskip}
        $\hfil\displaystyle{#1}\hfil$\crcr}}}           
\def\dt#1{{\buildrel {\hbox{\LARGE .}} \over {#1}}}     
\def\ddt#1{{\buildrel {\hbox{\LARGE .\kern-2pt.}} \over {#1}}}

\def\frac#1#2{{\textstyle{#1\over\vphantom2\smash{\raise.20ex
        \hbox{$\scriptstyle{#2}$}}}}}                   
\def\ha{\frac12}                                        
\def\sfrac#1#2{{\vphantom1\smash{\lower.5ex\hbox{\small$#1$}}\over
        \vphantom1\smash{\raise.4ex\hbox{\small$#2$}}}} 
\def\bfrac#1#2{{\vphantom1\smash{\lower.5ex\hbox{$#1$}}\over
        \vphantom1\smash{\raise.3ex\hbox{$#2$}}}}       
\def\afrac#1#2{{\vphantom1\smash{\lower.5ex\hbox{$#1$}}\over#2}}    

\def\boxes#1{
        \newcount\num
        \num=1
        \newdimen\downsy
        \downsy=-1.64ex
        \mskip-7.8mu
        \bo
        \loop
        \ifnum\num<#1
        \llap{\raise\num\downsy\hbox{$\bo$}}
        \advance\num by1
        \repeat}
\def\boxup#1#2{\newcount\numup
        \numup=#1
        \advance\numup by-1
        \newdimen\upsy
        \upsy=.82ex
        \mskip7.8mu
        \raise\numup\upsy\hbox{$#2$}}

\newskip\humongous \humongous=0pt plus 1000pt minus 1000pt
\def\caja{\mathsurround=0pt}

\newif\ifdtup
\def\panorama{\global\dtuptrue \openup2\jot \caja
        \everycr{\noalign{\ifdtup \global\dtupfalse
        \vskip-\lineskiplimit \vskip\normallineskiplimit
        \else \penalty\interdisplaylinepenalty \fi}}}
\def\li#1{\panorama \tabskip=\humongous                         
        \halign to\displaywidth{\hfil$\displaystyle{##}$
        \tabskip=0pt&$\displaystyle{{}##}$\hfil
        \tabskip=\humongous&\llap{$##$}\tabskip=0pt
        \crcr#1\crcr}}

\def\CMP{Commun. Math. Phys.}
\def\NP{Nucl. Phys. B}
\def\PL{Phys. Lett. }
\def\PR{Phys. Rev. Lett. }
\def\PRD{Phys. Rev. D}
\def\ref#1{$\sp{#1]}$}

\parskip=\medskipamount                 
\def\baselinestretch{1.2}       
\thicklines                         
\def\title#1#2#3#4{
\begin{document}
        {\hbox to\hsize{#4 \hfill Imperial-TP/ #3}}\par
        \begin{center}\vskip.5in minus.1in {\Large\bf #1}\\[.5in minus.2in]{#2}
        \vskip1.4in minus1.2in {\bf ABSTRACT}\\[.1in]\end{center}
        \begin{quotation}\par}
\def\author#1#2{#1\\[.1in]{\it #2}\\[.1in]}
\def\AM{Aleksandar Mikovi\'c\,
\footnote{E-mail address: A.MIKOVIC@IC.AC.UK}
\\[.1in] {\it Theoretical Physics Group, Blackett Laboratory, Imperial
College,\\ Prince Consort Road, London SW7 2BZ, U.K.}\\[.1in]}
\def\WS{W. Siegel\\[.1in] {\it Institute for Theoretical
        Physics,\\ State University of New York, Stony Brook, NY 11794-3840}
        \\[.1in]}
\def\endtitle{\par\end{quotation}\vskip3.5in minus2.3in\newpage}


\def\endabstract{\par\end{quotation}
        \renewcommand{\baselinestretch}{1.2}\small\normalsize}


\def\xpar{\par}                                         
\def\letterhead{
        \centerline{\large\sf IMPERIAL COLLEGE}
        \centerline{\sf Blackett Laboratory}
        \vskip-.07in
        \centerline{\sf Prince Consort Road, SW7 2BZ}
        \rightline{\scriptsize\sf Dr. Aleksandar Mikovi\'c}
        \vskip-.07in
        \rightline{\scriptsize\sf Tel: 071-589-5111/6983}
        \vskip-.07in
        \rightline{\scriptsize\sf E-mail: A.MIKOVIC@IC.AC.UK}}
\def\sig#1{{\leftskip=3.75in\parindent=0in\goodbreak\bigskip{Sincerely yours,}
\nobreak\vskip .7in{#1}\par}}


\def\ree#1#2#3{
        \hfuzz=35pt\hsize=5.5in\textwidth=5.5in
        \begin{document}
        \ttraggedright
        \par
        \noindent Referee report on Manuscript \##1\\
        Title: #2\\
        Authors: #3}


\def\start#1{\pagestyle{myheadings}\begin{document}\thispagestyle{myheadings}
        \setcounter{page}{#1}}


\catcode`@=11

\def\ps@myheadings{\def\@oddhead{\hbox{}\footnotesize\bf\rightmark \hfil
        \thepage}\def\@oddfoot{}\def\@evenhead{\footnotesize\bf
        \thepage\hfil\leftmark\hbox{}}\def\@evenfoot{}
        \def\sectionmark##1{}\def\subsectionmark##1{}
        \topmargin=-.35in\headheight=.17in\headsep=.35in}
\def\ps@acidheadings{\def\@oddhead{\hbox{}\rightmark\hbox{}}
        \def\@oddfoot{\rm\hfil\thepage\hfil}
        \def\@evenhead{\hbox{}\leftmark\hbox{}}\let\@evenfoot\@oddfoot
        \def\sectionmark##1{}\def\subsectionmark##1{}
        \topmargin=-.35in\headheight=.17in\headsep=.35in}

\catcode`@=12

\def\sect#1{\bigskip\medskip\goodbreak\noindent{\large\bf{#1}}\par\nobreak
        \medskip\markright{#1}}
\def\chsc#1#2{\phantom m\vskip.5in\noindent{\LARGE\bf{#1}}\par\vskip.75in
        \noindent{\large\bf{#2}}\par\medskip\markboth{#1}{#2}}
\def\Chsc#1#2#3#4{\phantom m\vskip.5in\noindent\halign{\LARGE\bf##&
        \LARGE\bf##\hfil\cr{#1}&{#2}\cr\noalign{\vskip8pt}&{#3}\cr}\par\vskip
        .75in\noindent{\large\bf{#4}}\par\medskip\markboth{{#1}{#2}{#3}}{#4}}
\def\chap#1{\phantom m\vskip.5in\noindent{\LARGE\bf{#1}}\par\vskip.75in
        \markboth{#1}{#1}}
\def\refs{\bigskip\medskip\goodbreak\noindent{\large\bf{REFERENCES}}\par
        \nobreak\bigskip\markboth{REFERENCES}{REFERENCES}
        \frenchspacing \parskip=0pt \renewcommand{\baselinestretch}{1}\small}
\def\unrefs{\normalsize \nonfrenchspacing \parskip=medskipamount}
\def\Item{\par\hang\textindent}
\def\Itemitem{\par\indent \hangindent2\parindent \textindent}
\def\makelabel#1{\hfil #1}
\def\topic{\par\noindent \hangafter1 \hangindent20pt}
\def\Topic{\par\noindent \hangafter1 \hangindent60pt}

\title{BLACK HOLES AND NONPERTURBATIVE CANONICAL 2D DILATON GRAVITY}
{\AM}{93-94/16}{February 1994}
We investigate nonperturbative canonical quantization of two dimensional
dilaton gravity theories with an
emphasis on the CGHS model. We use an approach where a canonical
transformation is constructed such that the constraints take a
quadratic form. The
required canonical transformation is obtained by using a method
based on the B\"acklund transformation from the Liouville theory. We
quantize dilaton gravity in terms of the new variables,
where it takes a form of a
bosonic string theory with background charges. Unitarity is then
established by going into a light-cone gauge. As a direct consequence,
black holes in this theory do not violate unitarity, and there is no
information loss. We argue that the information escapes during the
evaporation process. We also discuss the
implications of this quantization scheme for the quantum fate of real black
holes. The main conclusion is that black holes do not have to violate
quantum mechanics.

\endtitle

\sect{1. Introduction}

Since Hawking's discovery that black holes evaporate due to quantum
effects \cite{hawk}, the question of the quantum fate of a black hole has been
a subject of a lot of debate and speculation \cite{rev}. The reason
for this is a lack of a viable theory of quantum gravity, since
everybody agrees that at the end-point of the black hole evaporation
process quantum gravity effects will play a crucial role. In this
situation the best one can do is to study toy models of gravitational
collapse. The most realistic toy model studied so far is spherically
symmetric scalar field collapse in 4d \cite{unru}.
However, the resulting two-dimensional field theory is still to
complicated to be useful, so one has to resort to even simpler
models. Various toy models have been studied recently, including the spherical
dust shell collapse \cite{haj}, spherical dust cloud collapse
\cite{sdc}, and most notably, two-dimensional dilaton gravity
theories \cite{{rev},{gid}}. In particular, the CGHS model of 2d
dilaton gravity \cite{cghs} has attracted a lot of attention, since
it is classically exactly solvable,
and the solution describes formation of a 2d
black hole by massless scalar fields. Furthermore, there is a black
hole evaporation effect and the model represents a renormalisibile 2d field
theory \cite{cghs}.

These nice features of the CGHS model have raised a hope that already in
the one loop semi-classical approximation the effective equations of
motion will
be free of singular solutions \cite{cghs}. However, very soon it was
shown that singular solutions exist \cite{2dbh}, and furthermore
Hawking has argued that singularities exist in any semi-classical
approximation \cite{hawk2}. All this indicates that the 2d metric
has to be quantized if one wants to understand the quantum fate of the
black hole, and hence a nonperturbative quantization of the model is
necessary.

The nonperturbative approaches to 2d dilaton gravity which have been
studied so far are path-integral and canonical. The idea of the path-integral
approach is to perform the functional integral over the metric,
dilaton and matter fields exactly, and then to study the corresponding
effective action and the correlation functions (for a review and
references see \cite{{rev},{gid}}). Beside its own
difficulties, it is not clear how to construct the physical Hilbert
space within this approach, and how to address the corresponding
conceptual questions.

In the canonical approach \cite{{mik1},{mik2},{uch},{hir},{ver}},
the construction of the physical Hilbert
space is the primary goal, from which all other questions are
answered. This is achieved from the study of the constraints, which
can be derived by using the ADM method
\cite{mik1}. Alternatively, in the covariant quantization method
the space of the classical solutions defines the phase space and
the constraints are derived from the components of
the energy-momentum tensor \cite{{hir},{ver}}. The advantage of the
ADM method is its gauge independence, while
in the covariant approach a gauge has to be chosen from the beginning.
Either way, the idea is to simplify the constraints by constructing a
canonical transformation which makes the constraints quadratic in the
new canonical variables. Then one can derive the physical Hilbert
space by using the standard BRST techniques from the string theory
\cite{{mik1},{hir},{ver}}. Related to this is the issue of unitarity, which
was first discussed in \cite{mik2}. There it was demonstrateded that a
physical gauge exists where the unitarity can be easily proven.
Hence the black holes in this theory would not violate quantum mechanics.
Subsequently, a unitary S-matrix was constructed in the covariant
approach \cite{ver}, which was in accordance with the result of ref.
\cite{mik2}.

However, the Kac-Moody algebra method used in refs \cite{{mik1},{mik2}} to
construct the required
free-field canonical transformation is strictly valid only for the case of
chiral matter. On the other hand, although the free-field transformations of
refs \cite{{hir},{ver}} apply to the case of non-chiral matter, they
do not have the canonical form and are valid only in the conformal gauge.
In this paper we complete the program of canonical ADM
quantization of dilaton gravity. We construct the free-field canonical
transformation by using
methods based on the B\"acklund transformation from the Liouville theory.
Therefore we provide a strict proof of the unitarity of dilaton
gravity based on the ideas of ref. \cite{mik2}.

In section 2 we describe the canonical ADM formulation of 2d dilaton
gravity. In section 3 we introduce a Liouville dilaton gravity model,
as a preparatory study for the CGHS model. We construct the free field
canonical transformation after mapping the theory onto a usual
Liouville theory and then using the B\"acklund transformation. We also
find the B\"acklund transformation in the case when the Liouville energy
is not positive definite. Then in section 4 we discuss the CGHS model
by following the strategy from the Liouville model, i.e. we write the
fields of the
classical solution in the conformal gauge in terms of free fields and
then look for a simple relation between the corresponding canonical
variables. After obtaining such a relation, we show how the constraints
can be mapped into constraints of a string theory with background charges.
In section 5 we discuss the quantization of such a theory, and
describe the BRST
quantization in the compact case. In section 6 we describe the reduced
phase space quantization, which also applies to the non-compact case. We
find a physical gauge where all relevant dynamical variables can be
expressed in terms of independent canonical variables. The physical
Hilbert space is just a free-field Fock space for the matter, and the
corresponding Hamiltonian is a free-field one. It can be promoted into
a Hermitian operator, so that the quantum evolution is unitary.

We present our conclusions in sect. 7. We argue that the Hawking
radiation is present in spite of the trivial S-matrix. There is no
information loss, and we argue that the information escapes during the
evaporation. Implications for the real collapse are discussed and the
problem of singularity in quantum theory is addressed.

\sect{2. Canonical formulation}

Two-dimensional dilaton gravity theories of interest can be described
by an action
$$ \li{ S &= S_0 + S_m \cr
        S_0 &=  \int_{M} d^2 x \sqrt{-g} e^{-2\F}\left[ R +
\g (\nabla \F)^2 + U(\F ) \right]\cr
S_m &= -\ha\int_M d^2 x\sqrt{-g} \su_{i=1}^N (\nabla\f_i)^2 &(2.1)\cr}$$
where $\F$ and $\f_i$ are scalar fields, $\g$ is a constant, $g$, $R$
and $\nabla$ are
determinant, scalar curvature and covariant derivative respectively,
associated with a metric on the 2d manifold $M$. For our purposes we
will assume that $M= \S \times {\bf R}$, so that $\S = S^1$ (a circle) or
$\S = {\bf R}$ (a real line). We will
refer to these two cases as compact and non-compact respectively.
$S_0$ describes the
coupling of the dilaton $\F$ to the metric, while $S_m$ represents
conformally coupled scalar matter. Depending on the value of the
constant $\g$ and the form of the potential $U$, one can get various
dilaton gravity theories. For example, $\g =2$ and $U=\k e^{2\F}$
corresponds to the spherically symmetric Einstein-Hilbert action, while $\g
=4$ and $U=4\l^2$, where $\l$ is a constant, corresponds to the CGHS model.

Before explaining the canonical ADM formulation, we will briefly study
field redefinitions, in order to arrive at the simplest possible form of
the action. That in turn simplifies the constraints. Let $\j^2 =
e^{-2\F}$, then $S_0$ from the eq. (2.1) becomes
$$ S_0 =  \int_{M} d^2 x \sqrt{-g}\left[
\ha(\nabla \j)^2 +{1\over 2\g} R\j^2 + \tilde{U}(\j)\right]\quad,
\eqno(2.2)$$
where $\j$ has been rescaled into ${1\over\sqrt{2\g}}\j$ ($\g \ne 0$).
Then by performing a Russo-Tseytlin transformation \cite{tsey}
$$ \f = {1\over \g} \j^2 \quad,\quad \tilde{g}_{\m\n}=
e^{-2\r}g_{\m\n} \quad,\quad 2\r = {1\over \g} \j^2 - {\g\over 2}\ln\j
\eqno(2.3)$$
we get
$$ S_0 =  \int_{M} d^2 x \sqrt{-\tilde{g}}\left[
\ha(\tilde{\nabla} \f)^2 + \ha \tilde{R}\f +V(\f)\right]\quad, \eqno(2.4)$$
where $V(\f)= \tilde{U} e^{2\r}$. In the CGHS case $V =
\ha\l^2 e^\f$, and hence consider
$$ S_0 =  \int_{M} d^2 x \sqrt{-g}\left[
\ha(\nabla \f)^2 + \a R\f + \L e^{\b\f}\right]\quad, \eqno(2.5)$$
where $\a,\b$ and $\L$ are constants.
The action (2.5) represents a class of solvable dilaton gravity
theories, which can be seen by redefining the metric as $\tilde{g}_{\m\n}
= e^{\b\f}g_{\m\n} $ \cite{mik1}, so that
$$ S_0 =  \int_{M} d^2 x \sqrt{-\tilde{g}}\left[
\ha(1-2\a\b)(\tilde{\nabla} \f)^2 + \ha \tilde{R}\f + \L \right]\quad.
\eqno(2.6)$$
The action (2.6) is of the same form as the induced gravity action,
and one can use
the $SL(2,{\bf R})$ current algebra methods to construct the
free-field canonical transformation \cite{mik1}.
In the case when $\a\b =\ha$, the $SL(2,{\bf R})$
current algebra degenerates into an extended 2d Poincare current algebra,
and the analogous free-field construction exists \cite{uch}.
However, when $S_m$ is included, the current algebra method works
only for the case of chiral matter.

Therefore consider the following action
$$ S = \int_{M} d^2 x \sqrt{-g}\left[
{\g\over 2}(\nabla \f)^2 + \a R\f + V(\f)
-\ha\su_{i=1}^N (\nabla \f_i)^2 \right]\quad, \eqno(2.7)$$
where $\a$ and $\g$ are constants. Note that the field redefinitions
in eqs. (2.2-6) have always scaled the metric so that the form of
the matter action is unchanged, because of its conformal invariance.
Canonical reformulation
simplifies if we use the ADM parametrization of the metric
$$g_{\m\n}=\pmatrix{-{\cal N}^2 + gn^2 & gn \cr gn & g \cr} \quad,\eqno(2.8)$$
where $\cal N$ and $n$ are the laps function and the shift vector respectively,
while $g$ is a metric on $\S$. By defining the canonical momenta as
$$ p = {\pa \cl\over \pa \dt{g}}\quad,\quad \p = {\pa \cl\over \pa
\dt{\f}} \quad,\quad \p_i = {\pa \cl\over \pa \dt{\f_i}} \quad, \eqno(2.9)$$
where $\cl$ is the Lagrange density of (2.7) and dots stand for $t$
derivatives, the action becomes
$$ S= \int dt dx \left( p\dt{g} + \p\dt{\f} + \p^i\dt{\f_i} -
 N_0 G_0 - N_1 G\sb 1 \right) \quad, \eqno(2.10)$$
where $N_0 = {\cn\over\sqrt{g}}$ and $N_1 = n$.
The constraints $G_0$ and $G_1$ are given as
$$\li{G\sb 0 (x) =&   {\g\over2\a^2}(gp)^2
-{1\over\a}gp\p - {\g\over2}(\f^{\prime})^2 - gV(\f)
+ 2\a \sqrt{g}\left( {\f^{\prime}\over\sqrt{g}}\right)^{\prime}\cr & +
\ha \su_{i=1}^N (\p_i^2 + (\f_i^{\prime})^2) \cr
G\sb 1 (x) =& \p\f^{\prime} - 2p^{\prime}g - pg^{\prime}+
\su_{i=1}^N \p_i\f^{\prime}_i \quad,&(2.11)\cr}$$
where primes stand for $x$ derivatives.
The $G$'s form the canonical diffeomorphism algebra with respect to the
Poisson brackets. They also
generate the diffeomorphisms of $M$, such that $G_1$ generates the
diffeomorphisms of $\S$, while $G_0$ generates time translations of
$\S$ and hence it is called the Hamiltonian constraint.
A special feature of two dimensions is that
$$ T_{\pm} = \ha (G_0 \pm G_1) \eqno(2.12)$$
generate two commuting copies of the one-dimensional diffeomorphism
algebra. When $\S = S^1$ these two copies become two commuting
Virasoro algebras.

\sect{3. Liouville dilaton gravity}

As in the 4d canonical gravity, direct quantization of the constraints
(2.11) is difficult due to their non-polynomial dependence on the
canonical variables. One way around this problem is to follow the
strategy introduced by Ashtekar \cite{asht}, which is
to find new canonical variables such that the constraints become polynomial.
In the context of 2d gravity, this means that we will look for a canonical
transformation which makes the
constraints quadratic. We first study the model
$\a=1,\b\ne \ha$, which we are going to call Liouville dilaton
gravity, as a preparation for the CGHS model where $\a=1,\b=\ha$.
The Liouville dilaton gravity corresponds to
$U(\F)= 4\l^2 \exp [(\b-\ha)e^{2\F}]$.

Since the matter part of the constraints is already quadratic and
decoupled from the dilaton gravity sector, we need only to consider
pure dilaton gravity. After trivial rescaling of the dilaton,
eq. (2.6) becomes
$$ S_0 =  \int_{M} d^2 x \sqrt{-g}\left[
\ha(\nabla \f)^2 + \a R\f + \L \right]\quad,
\eqno(3.1)$$
where $2\a = {1\over\sqrt{1-2\b}}$. The constraints can be read off
from eq. (2.11)
$$\li{G\sb 0  =&   {1\over2\a^2}(gp)^2
-{1\over\a}gp\p - {1\over2}(\f^{\prime})^2 - g\L
+ 2\a \sqrt{g}\left( {\f^{\prime}\over\sqrt{g}}\right)^{\prime}\cr
G\sb 1  =& \p\f^{\prime} - 2p^{\prime}g - pg^{\prime} \quad.&(3.2)\cr}$$
By going into the conformal gauge $N_0 = 1$, $N_1 = 0$, and analyzing
the equations of motion for $g$ and $\f$, one can see that a
simplification occurs after the following canonical transformation
$$\li{\r = \a \ln g \quad,&\quad \p_{\r} = \frac1\a \, gp - \p \cr
      \j = \f + \a\ln g \quad,&\quad \p_{\j} = \p \quad.&(3.3)\cr}$$
The constraints then become
$$\li{G\sb 0  =& -\ha\p_\j^2 - \ha(\j^{\prime})^2 +
2\a\j^{\prime\prime} +\ha\p_\r^2 + \ha(\r^{\prime})^2 -
2\a\r^{\prime\prime} - \L \exp \left({\r\over\a}\right)  \cr
G\sb 1  =& \p_\j \j^{\prime}-2\a\p_\j^{\prime}
+\p_\r \r^{\prime} - 2\a\p_\r^{\prime} \quad.&(3.4)\cr}$$
Hence the new variables $\r$ and $\j$ are decoupled, and if $\L$ had been
zero our task would have been already accomplished. The
conformal factor $\r$ satisfies the Liouville equation in the
conformal gauge
$$ 4\pa_+\pa_- \r -{\L\over\a}\exp\left({\r\over\a}\right)=0 \quad,\eqno(3.5)$$
where $\pa_\pm = \pa/\pa x^\pm $ and $x^\pm = t \pm x$.
Since the $\r$-part of the constraints is the same as the
energy-momentum tensor of the Liouville theory, i.e.
$$G_0 =-\ha\p_\j^2 - \ha(\j^{\prime})^2 +
2\a\j^{\prime\prime} + T_{00}\quad,\quad G_1 = \p_\j
\j^{\prime}-2\a\p_\j^{\prime}  + T_{01} \quad,
\eqno(3.6) $$
where
$$ T_{00}=\ha\p_\r^2 + \ha(\r^{\prime})^2 - 2\a\r^{\prime\prime} -
\L \exp\left({\r\over\a}\right) \quad,\quad T_{01} = \p_\r
\r^{\prime}-2\a\p_\r^{\prime} \quad,
\eqno(3.7)$$
one can use the canonical form of the B\"acklund transformation \cite{back}
to make $T_{00}$ and $T_{01}$ quadratic
$$\p_\r = \o^{\prime} - \sqrt{-2\L} \exp {\r\over2\a} {\rm
sh}\left({\o\over2\a}\right)\quad,\quad \r^{\prime} = \p_\o -
\sqrt{-2\L} \exp {\r\over2\a} {\rm ch}\left({\o\over2\a}\right)\,\,.
\eqno(3.8) $$
When expressed in terms of the new variables $\j$ and $\o$, the
constraints become quadratic
$$\li{G\sb 0  =& -\ha\p_\j^2 - \ha(\j^{\prime})^2 +
2\a\j^{\prime\prime} +\ha\p_\o^2 + \ha(\o^{\prime})^2 -
2\a\p_\o^{\prime} \cr
G\sb 1  =& \p_\j \j^{\prime}-2\a\p_\j^{\prime}
+ \p_\o \o^{\prime} - 2\a\o^{\prime\prime}\quad, &(3.9)\cr}$$
which was our initial goal.

Note that the B\"acklund transformation (3.8) is defined only for
$\L<0$, which corresponds to positive definite $T_{00}$. When $\L>0$,
the analytic continuation of eq. (3.8) does not work, and
the corresponding expression can not be found in the Liouville
literature, since
it corresponds to the unphysical case of indefinite $T_{00}$. However,
the B\"acklund transformation still exists in that case, and can be found by
considering first the zero-mode case (no $x$ dependence) and by using the
method of generating function. Then it is
easy to see that the required transformation is
$$\p_\r = \o^{\prime} + \sqrt{2\L} \exp {\r\over2\a} {\rm
ch}\left({\o\over2\a}\right)\quad,\quad \r^{\prime} = \p_\o +
\sqrt{2\L} \exp {\r\over2\a} {\rm sh}\left({\o\over2\a}\right)\,\,,
\eqno(3.10)$$
which gives eq. (3.9) again.

\sect{4. CGHS model}

As we have shown in section 2, the dilaton gravity sector of the
CGHS model can be described by the action (2.6) with $\a= 1,\b=\ha$, so
that
$$ S_0 =  \int_{M} d^2 x \sqrt{-g}\left( R\f + \L \right)\quad,
\eqno(4.1)$$
where $\L = 4\l^2$. The corresponding constraints are then
$$\li{G\sb 0  =&  -gp\p - g\L
+ 2\sqrt{g}\left( {\f^{\prime}\over\sqrt{g}}\right)^{\prime}\cr
G\sb 1  =& \p\f^{\prime} - 2p^{\prime}g - pg^{\prime} \quad.&(4.2)\cr}$$
By introducing the conformal factor $\r$ as $g = e^\r$, so that
$p=e^{-\r}\p_\r$ we get
$$\li{G\sb 0  =& -\p_\r \p_\f  - \L e^\r + 2\f^{\prime\prime} -
\r^{\prime}\f^{\prime} \cr
G\sb 1  =& \p_\f \f^{\prime} + \p_\r \r^{\prime} -2\p_\r^{\prime}
\quad.&(4.3)\cr} $$

Now in order to find the required canonical transformation, we adopt
the strategy from the Liouville theory. We go into the conformal gauge
and study the structure of the classical solution in order to find the
analog of the B\"acklund transformation. In the conformal gauge $N_0
=1$ and $N_1 =0$ and the action (4.1) becomes
$$ S_0 = \int dt dx \left( \p_\r \dt{\r} + \p_\f \dt{\f} -
 G_0 \right) \quad. \eqno(4.4)$$
The equations of motion are then
$$\dt{\r} +\p_\f = 0 \quad,\quad \dt{\f} +\p_\r = 0 \eqno(4.5)$$
and
$$-\dt{\p_\r} - \f^{\prime\prime} + \L e^\r = 0 \quad,\quad
-\dt{\p_\f} - \r^{\prime\prime} = 0 \quad.\eqno(4.6)$$
By eliminating the momenta from eq. (4.5), eq. (4.6) becomes
$$4\pa_+\pa_- \f + \L e^\r = 0 \quad,\quad \pa_+\pa_- \r = 0
\quad,\eqno(4.7)$$
which can be solved as
$$e^\r = \pa_+ p(x^+)\pa_- m(x^-) \quad,\quad \f = a(x^+) + b(x^-) -
\l^2 p(x^+)m(x^-) \quad,\eqno(4.8)$$
where $p$ and $m$ are arbitrary functions, $a = a_0 + a_1 p$
and $b=b_0 + b_1 m$, where $a_\m$ and $b_\m$ are arbitrary constants.
When matter is present \cite{cghs}
$$ a = a_0 + a_1 p - \ha \int dx^+ \pa_+ p \int dx^+ {1\over \pa_+ p}
\su_i\pa_+ \f_i \pa_+ \f_i $$
$$ b = b_0 + b_1 m - \ha \int dx^- \pa_- m \int dx^- {1\over \pa_- m} \su_i
\pa_- \f_i \pa_- \f_i \quad.\eqno(4.9)$$

The task now is to define new fields in terms of $a,b,p$ and $m$, such
that there is an explicit relation between the old and the new
canonical variables, and that the constraints become quadratic. That
this is no simple task, one can see from the example of the Liouville
theory, where deriving the B\"acklund transformation from the classical
solution is not straightforward \cite{back}. The reason for this is
that the canonical transformation is a relation between fields and
their derivatives with respect to $t$ and $x$ coordinates,
while the classical
solution gives a relation between fields and their $x^+$ and $x^-$
derivatives. As a result, going from one to another description is not
trivial. For example, the constraints (4.3) are non-polynomial in
$\r$, but when expressed in the conformal gauge in terms of the
classical solution, one gets an quadratic expression
$$T_{\pm} = 2\pa_{\pm}\r \pa_{\pm}\f  - 2\pa_{\pm}^2\f \quad.\eqno(4.10)$$

In our case the obvious choice for the new variables is
$$\c = \ln \pa_+ p + \ln \pa_- m \quad,\quad \x = a + b \quad.\eqno(4.11)$$
Now we have to
look for a simple relation between $\f$, $\r$, $\c$ and $\x$ and their
$\pa_\pm$ derivatives which follows from (4.11) and (4.8).
The simplest two relations are
$$\pa_{\pm}\r \pa_{\pm}\f  - \pa_{\pm}^2\f = \pa_{\pm}\c \pa_{\pm}\x  -
\pa_{\pm}^2\x \quad,\eqno(4.12)$$
which is nothing else but the statement that $T_\pm$ stay quadratic
in the new variables.

By using the relations between the canonical
momenta and their velocities found in the equations of motion,
the two eqs in (4.12) can be rearranged to become
$$\li{ \p_\r \p_\f  + \L e^\r -2\f^{\prime\prime} +
\r^{\prime}\f^{\prime} =& \p_\c \p_\x  -2\x^{\prime\prime} +
\c^{\prime}\x^{\prime} \cr
 \p_\f \f^{\prime} + \p_\r \r^{\prime} -2\p_\r^{\prime} =& \p_\x \x^{\prime}
+ \p_\c \c^{\prime} -2\p_\c^{\prime}\quad.&(4.13)\cr} $$
Eq. (4.13) implies that $G_0$ and $G_1$ have become quadratic when
expressed in terms of the new variables. Also it can serve to
determine $(\f,\p_\f)$ and $(\r,\p_\r)$ in terms of $(\c,\p_\c)$ and
$(\x,\p_\x )$.

This can be done in the following way.
The form of the classical solution (4.8) and the relations between
the momenta and the velocities imply
$$\li{\r =\c \quad,&\quad \p_\r = \p_\c + \l^2 F(\c,\p_\x)\cr
 \f = \x + \l^2 G(\c,\p_\x) \quad,&\quad \p_\f = \p_\x \quad,&(4.14)\cr}$$
where $F$ and $G$ are functionals of $\c$ and $\p_\x$
to be determined. By inserting (4.14) into (4.13) we get
$$\li{ -F\p_\x + 4 e^\c + 2G^{\prime\prime} - \ha \c^{\prime}
G^{\prime} = & 0 \cr
-\p_\x G^{\prime} - F \c^{\prime} + 2 F^{\prime} =& 0
        \quad.&(4.15)\cr}$$
The system of eq. (4.15) can be rewritten as
$$G^{\prime} = -{1\over\p_\x}(F\c^{\prime} - 2F^{\prime}) \eqno(4.16)$$
and
$$ F^{\prime\prime} -
\left(\c^{\prime}+{\p^{\prime}_\x\over\p_\x}\right) F^{\prime} +
\frac14\left( (\c^{\prime})^2 - 2\c^{\prime\prime}
+2{\p^{\prime}_\x\over\p_\x}\c^{\prime}  - \p_\x^2 \right) F +
\p_\x e^\c = 0\quad. \eqno(4.17)$$

The eq. (4.17) is a second order ordinary differential equation
and hence a solution for $F$ exists. However, there is no explicit
expression for $F$ since the coefficients in eq. (4.17)
are arbitrary functions. In the zero-mode limit there is an explicit solution
$$ F= {4 e^\c\over \p_\x} \quad,\quad G=-{4 e^\c\over \p_\x^2} \quad,
\eqno(4.18)$$
and one can check then that the transformation (4.14) is canonical. This
is in contrast to the Liouville case, where the implicit relations (3.8)
give a first order ordinary differential equation
for $u = \exp(-{\r\over\a})$, and hence an
explicit expression for $(\r,\p_\r)$ as a function of $(\o,\p_\o)$
can be obtained. The lack of an explicit expression in the CGHS case
may seem like a big drawback, but, as we are going to show in the next
sections, there is enough useful information contained in the eq. (4.14) about
the free-field canonical transformation. Hence the eq. (4.14) can be
considered as the analog of the B\"acklund transformation from the
Liouville theory.

Before discussing the quantization, we perform a further canonical
transformation
$$ \c = {1\over\sqrt2}(\f_0 + \f_1) \quad,\quad
\x = {1\over\sqrt2}(\f_0 - \f_1) $$
$$ \p_\c ={1\over\sqrt2}( \p_0 + \p_1) \quad,\quad
\p_\x = {1\over\sqrt2}( \p_0 - \p_1)
\quad.\eqno(4.19)$$
The constraints then take the form
$$\li{ G_0 =& \ha ( \p_\m \p^\m + \f_\m^{\prime}\f^{\prime\m}) + Q^\m
\f^{\prime\prime}_\m \cr
G_1 =& \p^\m \f^{\prime}_\m + Q^\m
\p^{\prime}_\m \quad,&(4.20)\cr}$$
where
$\m = 0,1$, $\f_\m = (-\f_0 ,\f_1)$, $\f_\m = (\p_0 ,\p_1)$ and
$Q_\m = -\sqrt2 (1,1)$, while indices are raised with a metric $\h_{\m\n} =
{\rm diag} (-1,1)$. The same can be done in the Liouville case, where
$$ \j^{\prime} = \p_0 \quad,\quad \p_\j = \f_0^{\prime} $$
$$ \o = \f_1 \quad,\quad \p_\o = \p_1 \quad,\eqno(4.21)$$
and $Q_\m = -2\a (1,1)$. Note that in both cases $Q^2 =0$, which is
necessary in order for the expressions given by the eq. (4.20) to
satisfy the 2d canonical diffeomorphism Poisson bracket algebra.

\sect{5. Quantization}

In the canonical approach, there are two basic ways of
quantizing a constrained system

(1) quantize first and then solve the constraints (Dirac quantization),

(2) solve the constraints first and then quantize (reduced phase space
    (RPS) quantization).

\noindent The Dirac quantization, and its variations (Gupta-Bleuler
and BRST method), have an advantage over the RPS quantization
because the symmetries of the theory are manifest. On the
other hand, RPS quantization is easier to accomplish.
In our case, we are going to study both approaches. This is possible
because the constraints have the same form as the constraints of a $(
N+2)$-dimensional bosonic string with background charges, where many
quantization techniques have been developed.

In order to accomplish the Dirac quantization, it is useful to
introduce the left/right movers
$$ P_I^{\pm} = {1\over\sqrt2}(\pm\p_I + \f_I^{\prime}) \eqno (5.1)$$
which satisfy
$$ \{P_I^\pm (x),P_J^\pm (y)\} = \pm\h_{IJ} \d^{\prime}(x-y)\quad,\quad
\{P_I^+(x),P_J^-(y)\} = 0 \quad,\eqno(5.2)$$
where $I = \m,i$ and $\h_{IJ} = {\rm diag}. (-1,1,...,1)$.
Then the theory factorizes into two independent sectors, described by
$P^+$ and $P^-$ variables, with the respective constraints
$$T_{\pm} = \su_{I=1}^{N+2} \left( \ha P_I^{\pm} P_I^{\pm} + Q_I
P_I^{\prime\pm}\right) \quad.\eqno(5.3)$$
Hence it is sufficient to look for the physical Hilbert space in only
one sector, since the total physical Hilbert space will be a tensor product
of the left and the right sector.
Next we take $\S$ to be compact, because not much is known about the
representations of the 1d diffeomorphism algebra in the non-compact case.
This creates a problem, since
the black hole solutions strictly exist only in the non-compact case.
This problem is usually resolved
by putting the system into a large box, of length $L$, in the hope
that when $L \to \infty$ a non-compact case is recovered.

Now we make Fourier expansions
$$P_I (x) = {1\over\sqrt{L}}\left( p_I + \su_{n\ne 0} \a_n^I e^{in\p
x/L}  \right)\eqno(5.4)$$
where $p_i = 0$\footnote{This condition is necessary because $\f_i$
are matter fields and not the string coordinates, so that the Fock
space vacuum does not carry any momentum}. Then
$$\li{ T (x) = & {1\over L}\su_n L_n e^{in\p x/L} \cr
        L_n =& \ha\su_m \a_{n-m}^I \a_m^I + in Q_I \a_n^I \quad,&(5.5)\cr}$$
where $Q_i = 0$. The $L_n$'s are then promoted into operators acting on a
Fock space $\cf(\a^I_n)$ made out of the $\a_n$ modes in the standard way.
The $L_n$'s form a Virasoro algebra classically, but
in the quantum case there is an anomaly in the algebra, in the form of
the central extension term with the central charge $c =2+ N$
\footnote{Kuchar and Torre have shown that a new set of variables can
be found such that there is no anomaly in the diffeomorphism algebra.
However, the conformal symmetry then acquires the anomaly \cite{kt}}.
This type
of situation is best handled in the BRST formalism \cite{{mik1},{hir}}. One
enlarges the original Fock space $\cf (\a^I_n)$ by introducing a canonical
pair of ghost fields $(b,c)$, and constructs a nilpotent operator
$$ \hat{Q} = \su_n c_{-n}( L_n - a\d_{n,0}) +
\ha\su_{n,m}(n-m):c_{-n}c_{-m}b_{n+m} : \quad.\eqno(5.6)$$
The nilpotency of $\hat{Q}$ requires
$$ Q^I Q_I = -Q_0^2 + Q_1^2 = 2-N/12 \quad,\quad a = N/24 \quad,\eqno(5.7)$$
which is satisfied for $N=24$ since $Q^I Q_I = 0$.
The physical Hilbert space $\ch^*$ is then
determined as the cohomology of $\hat{Q}$
$$\ch^* = Ker\,\hat{Q}/Im\,\hat{Q} \quad.\eqno(5.8)$$

There is only a zero-ghost
sector in the cohomology, since the intercept $a\ne 0$. The physical
states satisfy
$$ ( L_0 - 1 )\Psi = 0 \quad,\quad L_n \Psi = 0 \quad n = 1,2
\quad,\eqno(5.10)$$
where $\Psi \in \cf (\a^I_n)$. The conditions (5.10) are well known in
the string theory, and they are satisfied by
the transverse oscilator states, corresponding to the $\a_n^i$ modes.
One can construct the corresponding observables (i.e. dynamical
variables which commute with the constraints) by using the DDF
construction \cite{ver}.

There are also discrete momentum states in the cohomology
\cite{hir}. However, their meaning in the context of dilaton
gravity is not clear since they do not have well defined scalar
product. As far as the continuous momentum states are concerned we
have been using the standard scalar product such that
$$ (\a_n)^{\dagger} = \a_{-n} \quad,\quad (L_n)^{\dagger} = L_{-n}
\quad.\eqno(5.11)$$

\sect{6. RPS Quantization}

One can arrive at the same results much more easily by using the RPS
approach. Since the constraint structure is the same as that of the
bosonic string, one can use the light-cone gauge to solve the
constraints \cite{gsw}. Another advantage of the RPS approach is that it also
works in the non-compact case, so that it avoids the problem of the
previous section.

The light-cone gauge is defined in terms of the $\f_\pm =
\f_0 \pm \f_1 $ variables, which in the CGHS case amounts to using the
$(\x,\c)$ variables. The standard light-cone gauge is
$$ \x = pt \quad,\quad \p_\c =- p \quad,\eqno(6.1)$$
where $p$ is $x$ independent. In the non-compact case $p$ is a
numerical constant (i.e. $p=1$), while in the compact case it is a
dynamical variable, representing the remaining global degree of
freedom of the dilaton gravity sector. By inserting the relations (6.1)
into the constraints we get
$$ \p_\x =- {1\over 2p}\su_{i=1}^N ( \p_i^2 + \f_i^{\prime 2} )\eqno(6.2)$$
and
$$\c^{\prime} ={1\over p}\su_{i=1}^N \p_i \f_i^{\prime}\quad. \eqno(6.3)$$
Hence the independent canonical variables are $(p,q)$ and
$(\p_i,\f_i)$, which agrees with the Dirac quantization result
that only the transverse mode states are physical.

The fact that the $G_0$ constraint can be put into the form (6.2) also
means that $\x$ is a time variable in the theory \cite{kuch}. Hence
a Hamiltonian can be associated
with the choice of time in (6.1), and it can be determined as
$$H = -\int_{\S} dx\, \p_\x = \ha \su_n (\a_{-n}^i\a_n^i +
\tilde{\a}_{-n}^i\tilde{\a}_n^i ) + c_0 \quad, \eqno(6.4)$$
where $\tilde{\a}$ are Fourier modes of $P^-$.
The constant $c_0$ is zero in the classical theory,
although it can have a quantum contribution due to the normal
ordering effects. In the non-compact case, the Hamiltonian can be
determined from
a surface term analysis \cite{bil}, but it is obvious from the equations of
motion for $\f_i$ that it is a free-field hamiltonian (6.4).

Unitarity of the theory follows from the fact
that the Hamiltonian
(6.4) can be promoted into a Hermitian operator acting on the
physical Hilbert space
$$\ch^* = \cf (\a^i_n) \otimes \cf (\tilde{\a}_n^i )= \cf(\a^i_k)
\quad,\eqno(6.5) $$
which is the usual free-field Fock space with $\a_k =
\a_n$ for $k>0$ and $\a_k = \tilde{\a}_n$ for $k<0$.
Therefore one has a unitary evolution described by a Schr\"odinger
equation
$$ i{\pa\over\pa t} \Psi (t) = \hat{H}\Psi(t) \quad,\eqno(6.6) $$
where $\Psi (t) \in \ch^*$, and hence no transitions from pure into
mixed states occur in this theory.

Although the gauge (6.1) is suitable for deriving the RPS Hamiltonian
and demonstrating the unitarity of the theory, it is not useful for
studying other relevant dynamical variables, like metric and curvature. The
CGHS conformal factor $\tilde{\r}$ is given by
$$e^{\tilde{\r}} = {e^\r\over \f} \eqno(6.7)$$
and by using the eq. (4.14), $\tilde{\r}$ can be expressed in the gauge
(6.1). However,
that expression is not very useful since we do not know the explicit
expression for $F$. On the other hand, when chosing the light-cone
gauge, instead of using $(\f_-,\p_+)$ variables one can use $(\f_+,\p_-)$
variables, or equivalently $(\c,\p_\x)$. The advantage of doing this is
evident from the eq. (4.14), since it implies $\c = \r$ and $\p_\x =
\p_\f$. Then
from the constraint equations one can obtain an explicit expression for $\f$.
Hence the gauge
$$ \c =0  \quad,\quad \p_\x = 0 \quad,\eqno(6.8)$$
is equivalent to
$$\r =0 \quad,\quad \p_\f = 0 \quad,\eqno(6.9)$$
so that the constraint equations become
$$G_0 = -4\l^2 + 2 \f^{\prime\prime} + \ha \su_i (\p_i^2 + \f_i^{\prime 2}) =
0 \eqno(6.10)$$
and
$$G_1 = -2\p_\r^{\prime} + \su_i\p_i\f_i^{\prime} = 0\quad. \eqno(6.11)$$
 From eq. (6.10) we get
$$ \f = A + B x + \l^2 x^2 -\frac14 \int dx \int dx \su_i (\p_i^2 +
\f_i^{\prime 2}) \quad.\eqno(6.12)$$
When compared to the solution (4.8-9), we see that
the gauge (6.8) corresponds to $p = x^+$ and $m = x^-$ and
$$ A = a_0 + b_0 - \l^2 t^2 \quad,\quad B = -a_1=b_1\quad.
\eqno(6.13)$$
This illustrates the fact that chosing a canonical gauge is the same as chosing
coordinates on $M$.
The corresponding Hamiltonian is again the free-field Hamiltonian (6.4).

\sect{7. Conclusions}

The immediate conclusion is that a unitary quantum theory of 2d dilaton
gravity can be constructed. In turn that implies that the 2d black holes
in such a theory do not destroy information, and a unitary S-matrix exists.
Furthermore, the S-matrix is trivial
since the matter is described by a free-field Hamiltonian.
This makes one suspicious whether black holes in such a theory have
semi-classical properties we wanted to study in the first place, most
notably do they evaporate. That black hole evaporation can occur in the
theory can be seen from the following argument \cite{mik2}.

Let $\Psi_0$ be a physical state at $t=0$ such that
$$ <\Psi_0|\hat{g}(x)|\Psi_0> \quad,\quad
<\Psi_0|\hat{R}(x)|\Psi_0> \eqno(7.1)$$
are regular functions for every $x\in \S$. $g(x)$ is a spatial metric,
given by eq. (6.7), and in the gauge (6.8) $\f$ is given by eq.
(6.12). $R(x)$ is the scalar curvature, which can be also expressed in
terms of the free fields, via the formula $R =
e^{-\tilde{\r}}\pa_+\pa_-  \tilde{\r}$.
Promoting these variables into well defined Hermitian operators is not
an easy task, but the results of the covariant approach
\cite{jap} imply that it can be done.
The time evolution of $\Psi_0$ is then given by
$$\Psi(t) = e^{-i\hat{H}t}\Psi_0 \quad,\eqno(7.2)$$
and when the apparent horizon forms in the effective metric
$<\Psi(t)|\hat{g}(x)|\Psi(t)>$, one can split the modes of $\f_i$ into
these which are inside the horizon and those which are outside the horizon.
Then a density matrix $\hat{\r}(t)$ can be associated with $\Psi (t)$
by tracing out the states corresponding to the modes inside the horizon.
Given the density matrix $\hat{\r}(t)$, one could find out when it takes
approximately the thermal form
$$ \hat{\r} \approx {1\over Z}e^{-\b \hat{H}_+} \quad,\eqno(7.3)$$
where $H_+$ is the Hamiltonian of the modes outside the horizon.
More precisely, let $\Psi_0$ be a semiclassical state such that
the effective quantum metric
$$g_{eff} (t,x) = <\Psi_0|\hat{g}_H (t,x)|\Psi_0> = g_{b.h.}(t,x)\left( 1 +
\e \d g_1 (t,x) + \e^2 \d g_2 (t,x) + ... \right) \quad,\eqno(7.4)$$
where $\e$ is a small dimensionless parameter constructed from $\l$, Planck's
constant and parameters of the matter state $\Psi_0$.
$\hat{g}_H (t,x) = e^{iHt}\hat{g}(x)e^{-iHt}$, while $g_{b.h.}$ is a black hole
metric corresponding to the matter distribution described by the semiclassical
state $\Psi_0$. The corrections $\d g_n$ describe the backreaction effects of
the quantum matter on the classical metric, and can be calculated from the
identity
$$\hat{g}_H = (-\l^2 x^+ x^- - \hat F )^{-1} = (-\l^2 x^+ x^- - <\hat{F}>)^{-1}
(1 - \hat{\d F})^{-1} \eqno(7.5)$$
where
$$ \hat{\d F} = (-\l^2 x^+ x^- - <\hat{F}>)^{-1}(\hat{F}-<\hat{F}>)=
g_{b.h.}(\hat{F}-<\hat{F}>)\quad,\eqno(7.6) $$
so that
$$ \e^n \d g_n = <\hat{\d F}^n>\quad.\eqno(7.7)$$
All expectation values are with respect to $\Psi_0$, and
$$ \hat{F}= \ha \int dx^+ \int dx^+ :\pa_+\hat{\f}_i\pa_+ \hat{\f}_i: +
\ha \int dx^- \int dx^- :\pa_- \hat{\f}_i \pa_- \hat{\f}_i: \quad,\eqno(7.8)$$
where the normal ordering is with respect to the in vacuum $|0_{in}>$, which
is defined with respect to the $g_{b.h.}$ metric at $t=-\infty$.
In the semiclassical
approximation the backreaction effects are neglected, and therefore one
gets a quantum free field propagation on the black hole background.
Furthermore if $\Psi_0$ is chosen such that $\Psi (t)$ is close to the
$|0_{in}>$ for late times, then by the standard argument \cite{gn}, the density
matrix will be thermal for late times with the temperature equal to the
Hawking temperature
$$ T = {\l\over 2\p} \quad.\eqno(7.9)$$
This program will involve some non-trivial calculations, and still has
to be tested, but we do not see anything in principle which could
spoil this scenario.

If we accept the argument that the theory has a correct semiclassical
limit together with the fact that the theory is unitary, we are then in
position to say something about the information loss problem. Since
the matter comes in and out of the black hole without any hindrance,
we can say that there are no remnants, because they would correspond
to a formation of a bound state. Hence the information which falls in
has to escape during the evaporation process. This scenario is usually
criticized
on the grounds that it violates locality and causality\cite{{rev},{gid}}.
However, in a quantum theory, non-locality is unavoidable
since quantum mechanics is not local, which was
confirmed by numerous EPR-type experiments. Moreover, as demonstrated
in the case of the EPR experiment \cite{shim}, this non-locality may
be such that information cannot be transmitted faster than light.
Another point is that the usual notion of locality is defined for a
field theory on a fixed spacetime background, while in quantum
gravity the metric
is a hermitian operator, so that the definite spacetime
background only emerges for special states.

One can also study the unitarity of the theory
in the S-matrix formalism \cite{ver}. However, in order to
obtain a non-trivial scattering matrix, the authors of \cite{ver} had
to modify the theory by introducing a reflecting boundary
condition. Such a theory is locally the same as 2d dilaton gravity,
and the S-matrix is unitary, which is in agreement with our results.
It is interesting that the studies of other toy models of gravitational
collapse \cite{{haj},{sdc}} have given the same result, i.e. that a
unitary quantum
theory of such models exists. This is a very good sign for believing
in the existence of a unitary theory of a real collapse, because if we
had not been able to make the toy models unitary, than it would not have
been any hope for the full theory. On the other hand, demonstrating the
unitarity of a real collapse will be an extraordinary task. Even in
the spherically symmetric case, the equations of motion
are not integrable, which
means that the technique of constructing free-field canonical
transformation is not going to work. Most probably one would have to
adapt techniques developed in the context of full general relativity
with matter,
like loop variables \cite{rov}. The problem of time in quantum gravity
may seem like an immediate obstacle for a unitary theory, but this is
really the problem of quantum cosmology, while in the gravitational collapse
we are dealing with the asymptotically flat boundary conditions, hence
a global time is always available.

We must also stress that in the framework of canonical quantum gravity
one deals only with a single universe, and spacetime topology changes
where a baby universe or a wormhole is created cannot be addressed by the
formalism. Such processes can be used to explain the information loss,
but the problem with this is that they require a third
quantized theory of gravity, which is even less understood than the
canonical quantum gravity.

Beside the question of unitarity of gravitational collapse, the
question of the fate of the classical singularity in the quantum
theory requires
an explanation. Although we have managed to find observables
in our toy model which are well defined at the singularity,
there will be
other observables, those associated with the scalar curvature, which will
not be well defined at the singularity. One way of resolving this problem
is to study an effective scalar curvature \cite{mik2}
$$ R_{eff}(x,t) =
<\Psi_0|e^{i\hat{H}t}\hat{R}(x)e^{-i\hat{H}t}|\Psi_0>\quad. \eqno(7.10) $$
If it stays a regular function for every $x$ and $t\ge 0$ and for every
$\Psi_0$  that satisfies
the conditions of eq. (7.1), then we could say that the singularity
has been removed from the quantum theory. A very similar idea has been
proposed in \cite{smo}. However, there is no
a priori reason for something like this to happen, and the result will
depend on the dynamics of the theory. Note that such approach
has been already tried in the context of mini-superspace
cosmological models \cite{ashetal}. In analogy with the 2d dilaton
gravity, a canonical transformation was constructed which maps the
Hamiltonian constraint into a quadratic form. As a consequence the
quantum evolution
is unitary, but the classical singularity is not removed,
since the expectation value of the Weyl curvature scalar is
infinite at $t=0$ for every physical state.

One may hope that something like this will not
happen in the case of dilaton gravity models, because of the extra
smearing due to
presence of a spatial coordinate. However, judging from the results of
the covariant approach analysis \cite{jap}, it is very likely that the
effective scalar curvature (7.10) will not stay finite for all
regular initial states. Still, it is not
clear what does this mean, since one can invoke the argument that
the initial state has evolved into a state which does not have a
spacetime interpretation. Clearly, further studies are necessary in
order to clarify this issue.

\end{document}